\def\gsim{\mathrel{\rlap{\lower.6ex\hbox{$\sim$}}\raise.35ex\hbox{$>$}}}
\newcommand {\be}{\begin{equation}}
\newcommand {\ee}{\end{equation}}

\newcommand {\bea}{\begin{eqnarray}}
\newcommand {\eea}{\end{eqnarray}}
\documentstyle[preprint,aps,eqsecnum]{revtex}

\begin{document}
\preprint{September 1994}
\title
{Geometrical Approach to Bosonization
of $D>1$ Dimensional (Non) Fermi Liquids}

\author{D.V. Khveshchenko}
\address
{Department of Physics, Princeton University,\\ Princeton, NJ 08544\\
and\\
Landau Institute for Theoretical Physics,\\
2, st.Kosygina, 117940, Moscow, Russia}
\maketitle

\begin{abstract}
\noindent
We discuss an approach
to higher dimensional bosonization of interacting fermions
based on a picture
of fluctuating Fermi surface. Compared with the linearized
"constructive" approach developed in Refs.
\cite{H},\cite{HM},\cite{FC}
this method allows an account of the Fermi
surface curvature due to
nongaussian  terms in the bosonized Lagrangian.
On the basis of this description we propose a procedure of
 calculating density response functions
beyond the random phase approximation.
We also formulate a bosonic theory
 of the compressible metal-like
state at half filled lowest Landau level and check that
in gaussian approximation it reproduces RPA results
of the gauge theory by Halperin, Lee, and Read.

\end{abstract}
\pagebreak

\section {Introduction}
Recently it has been a lot of interest in possible scenaria of
a breakdown of
Landau-Fermi liquid behavior in two- and three-dimensional strongly
correlated systems. In particular, it was conjectured that
it might happen in presence of long-ranged density-density
or,
even more likely, current-current interactions \cite{A1},\cite{S},\cite{BW},
\cite{KHR}.
In particular,
a system of
nonrelativistic fermions coupled with an abelian gauge field was argued
to  present
such an example \cite{R}.

However a systematic investigation of these exciting
possibilities still remains to be carried out. One essential reason is
a lack of an adequate formalism capable of giving a proper description of
singular interactions which may result
to non-Fermi liquid states. Although standard calculations
in low order perturbation theory can, in principle, reveal features
signalling about a non-Fermi liquid regime one certainly needs a more advanced
technique to study a suspicious problem in greater details.

Among possible improvements a renormalization group wethod was recently
proposed and tested in the case of short-range interactions which do not
destroy Fermi liquid in $D>1$ (except for an instability in
the Cooper channel) \cite{Sh}.

Another promising development was provided by the method of higher dimensional
bosonization. It was first discussed in \cite{L} in a manner close
to the more recent Anderson's picture of "Tomographic projection" \cite{A2}
where a D-dimensional space
is considered as a set of essentially
uncoupled one-dimensional "rays".
In its original form the method of the Ref.\cite{L} was basically intended
 to reproduce fermion algebra and correlations along a given radial direction
in momentum space. As an output
a bosonic representation of free fermion
correlation functions was first obtained  \cite{L}.

More recently it was proposed
by Haldane \cite{H} and then elaborated in
\cite{HM},\cite{FC} how to treat couplings between different Fermi points
in the bosonized theory. This generalization involves a construction of
effective bosonic
variables as sums over squat boxes ("patches") instead of radial rays
in  momentum space.
In the framework of this description
main results of the Landau-Fermi liquid theory were
recovered
\cite{HM},\cite{FC}.
However when applied to less familiar problems
such as a problem of two-dimensional nonrelativistic fermions
coupled with a gauge field the method leads
to results of uncertain status \cite{M}. Namely, the authors
of \cite{AIM} who studied the same problem by using a self-consistent
diagrammatic approach argued that the results obtained in  \cite{M}
can be only valid in the unphysical limit of
a zero number of fermion species.

An obvious shortcoming of the "constructive" approach
to higher dimensional bosonization
 is that it does not provide a proper account of Fermi surface
curvature. Technically, it means that within this scheme
one can not treat properly transferred momenta tangential
to the Fermi surface while these become more and more important as
one gets closer to it.

For instance, if it turns out that at  given energy transfer
$\omega$ an interaction vertex determines an average
tangential transferred momentum to be $q_t >> \omega^{1/2}$ then the term
$\sim q^2$ which is due to a finite Fermi surface curvature
can not be neglected in typical denominators
$\omega -{\vec v}_F {\vec q}+ {1\over 2m}q^2$ which appear in
integral expressions for Green functions
obtained via bosonization \cite{HM},\cite{FC} (see also \cite{CCM}).

In particular, omitting $q_t$ in the problem of fermions with gauge
interactions one can never get to the regime where the Migdal theorem stating
an irrelevance of vertex corrections holds. On the contrary, it was argued in
\cite{Pol},\cite{AIM} that in the case of
the $D>2$ gauge problem with $q_t \sim \omega^{1/3}$ the Migdal theorem
always holds (in the 2D case it can be only valid in the limit of infinite
number of fermion species \cite{AIM}).

We mention, in passing, that the general method of eikonal
applied to the 2D gauge
problem in \cite{KS} is, in fact, not plagued with
this flaw  and provides
a natural account of $\sim q^2$ terms.
However in \cite{KS} these terms were deliberately omitted to obtain
an explicit form of the one-particle Green function.
Thus the formula for the one-particle Green function obtained in \cite{KS}
ceases to be valid in the very vicinity of the Fermi surface. On the other
hand, including abovementioned terms $\sim q^2$ one
ends up with an expression which agrees with the results of \cite{AIM}
in both regimes when either $\epsilon$ or $v_F (p-p_F)$ is much greater than
another.

Postponing a thorough revision of the eikonal results of \cite{KS}
until another paper we shall
present here
a "geometrical" approach to $D>1$ bosonization which allows a
systematic account of tangential components of transferred momenta $q_t$.
We shall also illustrate how the  method works in the case of 2D
density-density as well as current-density interactions.

\section{Nonlinear bosonization
 of interacting fermions in external fields}

Talking about bosonization one actually means a procedure of
calculating
(gauge invariant) response functions using functional integrals
in terms of some bosonic variables.
Remarkably, dealing with this problem one can avoid a subtle question about
an explicit construction of the fermion operator in terms of those bosons.
Although the corresponding formula
was repeatedly conjectured in
\cite{L}, \cite{H},\cite{HM},\cite{FC},
a necessity to supplement a naive D-dimensional counterpart
of the 1D relation
$\psi\sim\exp i\phi$ by a complicated ordering operator
makes this representation hardly useful in practice.
Moreover a calculation of the fermion Green function itself does not
provide much physical insight in  cases where some
 gauge symmetry is involved.

Nevertheless,
a systematic approach to higher dimensional bosonization can be developed
in the framework of the general method of "coadjoint orbit quantization"
\cite{coherent}.
Adopting this general procedure to the case of interacting fermions
one may choose a basis of coherent
states
\be
|\{g({\vec p}, {\vec q})\}>={\hat g}|vac>=
\exp (i\int d{\vec p}d{\vec q} g({\vec p}, {\vec q}){\hat
n}_{\vec p}(\vec q))|vac>
\ee
created by elements ${\hat g}$
of the infinite group $G$ from some vacuum state $|vac>$.
The group $G$
is generated by  fermion bilinear operators
${\hat n}_{\vec p}(\vec q)=\psi^{\dag}({\vec p}+{\vec q})\psi({\vec p})$
 obeying the algebra (referred in 1D case as $W_\infty$)
\be
[{\hat n}_{\vec p}(\vec q), {\hat n}_{\vec p^{\prime}}(\vec q^{\prime})]=
\delta({\vec p}-{\vec p}^{\prime}-{\vec q}^{\prime})
{\hat n}_{{\vec p}^{\prime}}({\vec q}+{\vec q}^{\prime})
-
\delta({\vec p}^{\prime}-{\vec p}-{\vec q})
{\hat n}_{{\vec p}}({\vec q}+{\vec q}^{\prime})
\ee
The orbit of the group $G$ associated with some reference
ground state consists of elements
\be
Q=
{\hat g}|vac><vac|{\hat g}^{-1}
\ee
An expectation value of the operator ${\hat A}$ taken over a coherent
state (2.1) can be expressed in terms of $Q$ as follows
\be
<{g}|{\hat A}|g>=
tr \{Q{\hat A}\}
\ee
where the trace stands for the integral over phase space
$tr=\int d{\vec x}d{\vec q}$ and ${\vec x}$ is a Fourier transform of
$\vec p$.

Under quantization by means of
  the functional integral
the bosonic variable
\be
w({\vec p}, {\vec q})=<{g}|{\hat n}_{\vec p}(\vec q)|{g}>=
tr\{Q{\hat n}_{\vec p}(\vec q)\}
\ee
 parametrising the orbit element (2.3) becomes a quantum field
which
can be identified with the partial Fourier transform of
the quantum Wigner distribution
function ${\tilde w}({\vec p}, {\vec r},t)$ (phase space density).

To write down the bosonic Lagrangian in terms of $w({\vec p}, {\vec q})$
one has to find analogue of the term $p{\dot q}$. It turns out
that it can be only  written
in terms of a cocycle
\cite{coherent} by introducing a fictitious variable
$u$ such that $Q({\vec p}, {\vec q},u=0)=Q({\vec p}, {\vec q})$.
Then the Lagrangian acquires the form
\bea
L=<g|i\partial_{t} - H|g>=\\ \nonumber
=i\int_{0}^{\infty} du tr (Q\{ {\partial_{u}Q},
{\partial_{t} Q} \}_{MB}) - tr(HQ)
\eea
where $\{A,B\}_{MB}$ stands for the so-called Moyal bracket
\be
\{A,B\}_{MB}={2\over \hbar}
\sin{\hbar\over 2}(\partial_{x}\partial_{q^{\prime}}-
\partial_{q}\partial_{x^{\prime}})A(x,q)B(x^{\prime},q^{\prime})
|_{x=x^{\prime},q=q^{\prime}}
\ee
which amounts to the Poisson one in the semiclassical limit $\hbar\rightarrow
0$.

The functional integral for the theory (2.6) written
in terms of $w({\vec p}, {\vec q})$ provides an exact
bosonization of the original fermion problem even in the case of a nonlinear
bare fermion spectrum and/or nonlocal interactions in any dimension
\cite{ind},\cite{K}.
However because of the overcompleteness of the basis of coherent states
of ${\cal N}$ particles the
variables $w({\vec p}, {\vec q})$ have to be subjected
to additional constraints
\be
Q^2=Q, ~~~~~~~ tr Q={\cal N}
\ee
which make things too complicated.
Nevertheless this description can be used for a systematic derivation of
corrections due to spectrum nonlinearity and/or nonlocality of interactions.
In the lowest order in gradients
the method is essentially equivalent
to the "constructive" bosonization approach \cite{H},\cite{HM},\cite{FC}
where the right hand side of the commutation relations (2.1) is replaced by
a c-number
\be
[{\hat n}_{\vec p}({\vec q})
, {\hat n}_{{\vec p}^{\prime}}({\vec q}^{\prime})]\approx
{1\over 2\pi}\delta({\vec p}-{\vec p}^{\prime})
\delta({\vec q}+{\vec q}^{\prime})
({\vec q}{\vec \nabla}_{\vec p})n_{\vec p}^{(0)}
\ee
where $n_{\vec p}^{(0)}=\theta (k_F -p)$ is the bare
Fermi distribution function.

 Depending on details of interaction,
the approximate commutation relations (2.9)
between ${\hat n}_{\vec p}({\vec q})$
may become asymptotically correct in a sense
of low energy, small angle scattering matrix elements of both sides
of (2.9).
However the
constructive method
encounters the problem of an artificial low-energy cutoff $\Lambda <<k_F$
which
appears
in the construction of oscillator-like bosonic variables
$a_{\vec n}({\vec q})={1\over {\sqrt{({\vec n}{\vec q})}}}
\sum_{|p_F {\vec n}-{\vec p}|<\Lambda}
\psi^{\dag}({\vec p}+{\vec q})\psi({\vec p})$ defined as sums over
 patches of  size $\Lambda$ on the Fermi surface.

Presumably, the whole procedure can be only trusted if
this cutoff does not enter
physical observables.

On the contrary, such a problem  simply does not occur in the formally
exact method which we described above.  Notice, that it can be thought  as
a hydrodynamical
field theory which generalises the phenomenological Landau theory
including
collisions between quasiparticles.
However to make this approach  really working
one needs a more convenient parametrization
of the phase space density in terms of some bosonic variables which
resolve the
constraints (2.8) imposed on the Wigner functions $w({\vec p}, {\vec q})$.

It seems natural enough to formulate
a purely bosonic description of interacting fermions in terms of the vector
variable ${\vec k}_F ({\vec r}, t, {\hat s})$ which traces the shape of the
$D-1$-dimensional Fermi
surface parametrised by coordinates ${\hat s}=s_1 , ...,s_{D-1}$
and varying from one space-time point to another.

It is also consistent with the  conjecture made
in the course of the previous work \cite{H},\cite{FC}
that all relevant low-energy
processes can be described as
fluctuations of the Fermi
surface viewed as an extended dynamical object.

Notice that such kind of description
had already appeared in the theory of edge states of Quantum Hall droplets
where  one treats edge states as capillary waves on the droplet boundary
which plays a role of Fermi surface in the case of a strong magnetic field
\cite{IKS}.
It turns out that this  description naturally occurs
in the framework of the theory (2.6)
if one considers Fermi surfaces with a sharp boundary
and approximates ${\tilde w}({\vec p},{\vec r},t)$ by a support function
\be
{\tilde w}({\vec p},{\vec r},t)\approx
\theta(|{\vec k}_F ({\vec r}, t, {\hat s})|-p)
\ee
which becomes appropriate in the longwavelength limit.

One may also understand ${\vec k}_F ({\vec r}, t, {\hat s})$
as a sum ${\vec k}_F ({\vec r}, t, {\hat s})=
{\vec k}_{F_0}({\hat s})+{\vec \nabla}\phi({\vec r}, t, {\hat s})$
where the first term  corresponds to some reference shape of the
Fermi surface (not necessarily circular) while the second one describes
fluctuations around it.

Physically,
the field $\phi({\vec r}, t, {\hat s})$ has a  meaning of a phase
of a wave function of
wave packet created at some space-time
point $x=({\vec r}, t)$ with a momentum in the direction ${\hat s}$.
Then the variable $\phi$ is defined modulo
$2\pi$ which opens a possibility of nontrivial winding numbers
along noncontractable contours on the Fermi surface (if any).

According to this physical interpretation
we  consider $\phi$  as a primary field with
fundamental equal-time commutation relations (compare with \cite{IKS})
\be
[\phi({\vec r}, {\hat s}),
({\vec n}{\vec \nabla})\phi({\vec r}^{\prime}, {\hat s}^{\prime})]=
2\pi i\delta({\hat s}-{\hat s}^{\prime})
\delta({\vec r}-{\vec r}^{\prime})
\ee
where
${n}_{\mu}(s) ={1\over {|\partial_{s}{\vec  k}|^{D-1}}}
\epsilon_{\mu\nu_{1} ...\nu_{D-1}}{\partial_{s_1} k_{\nu_1}}\times ...
\times{\partial_{s_{D-1}} k_{\nu_{D-1}}}$
is a normal unit vector to the reference Fermi surface ${\vec k}_{F}
={\vec k}_{F_0}
(s)$.

Notice that the commutation relations (2.11) are not canonical since the
momentum variable conjugated to $\phi({\vec r}, {\hat s})$
can be expressed as its gradient. It is consistent with the idea
of a chiral nature of the boson field
$\phi({\vec r}, {\hat s})$
which obeys the first order equation of motion \cite{H},\cite{HM},\cite{FC}.

To proceed further one has to express a local density operator $\rho
({\vec r},t)$ in terms of ${\vec k}_F ({\vec r}, t, {\hat s})$.
Only for the sake of notational simplicity we shall concentrate
on the 2D spinless case.

The wanted relation can be readily found on purely
geometrical grounds since in our approximation the
total density is nothing but
the area in momentum space enclosed by the curve ${\vec k}_F =
{\vec k}_F({s})$ parametrised by a $2\pi$-periodic variable $s$:
\be
\rho={1\over 2}\oint {\partial_s {\vec k}_F}\times
{\vec k}_F{ds\over (2\pi)^2}
\ee

Subtracting the constant term $\partial_s {\vec k}_{F_0}\times
{\vec k}_{F_0}$ from the integrand in (2.12)  we can write the remainder as
\be
\rho_s =
{1\over 2\pi}(|\partial_s {\vec k}_{F_{0}}|({\vec n}_s{\vec \nabla})
\phi+{1\over 2}{\vec \nabla}\partial_s \phi
\times {\vec \nabla}\phi)
\ee
so that $\rho ({\vec r})=\rho_0 +\oint {ds\over 2\pi}\rho_s ({\vec r})$
where $\rho_0 =
{1\over 2}\oint{ds\over 2\pi}\partial_s {\vec k}_{F_0}\times
{\vec k}_{F_0}$ is an
average density.

The quantity $\rho_s$  associated with a point $s$ on the Fermi surface
 plays a role similar to the oscillator-like partial density operator
$\sqrt{{\vec q}{\vec n}}a_{\vec n}({\vec q})$ introduced
in \cite{H},\cite{HM},\cite{FC}.
 It also follows from (2.10) that
$\delta w({\vec k}_{F_0}, {\vec q})
\approx\delta\rho_s ({\vec q}){\partial n^{0}(p)\over {\partial \epsilon_p}}$.

Apparently, in presence of some gauge vector potential $\vec A$ our
${\vec k}_F$ is modified as ${\vec k}_F\rightarrow
{\vec k}_{F_0}+{\vec \nabla}\phi-{\vec A}$.
All time derivatives
also become covariant: ${\partial_t \phi}\rightarrow
D_0 \phi=
{\partial_t \phi}-A_0$. Concerning derivatives with respect
to the parameter $s$ labeling points of
 the Fermi surface one can also include
a new component of the gauge field
$A_s$ corresponding to a reparametrization invariance of the
Fermi surface. However in the following we shall leave this interesting
possibility apart and simply fix the gauge by
putting $A_s=0$.

One can also see
that the partial density $\rho_s$ is gauge non-invariant quantity
while the total density $\rho$
does not depend on the gauge field $A_\mu$
since ${\partial_s {\vec A}}=0$ and
$\oint {\partial_s {\vec k}}{ds\over 2\pi}=0$.

The formula (2.12) allows a straightforward generalization onto the case
of arbitrary dimension $D$:  $\rho={1\over D}\oint
\epsilon_{\mu_1 ...\mu_D}{\partial_{s_1} k_{\mu_1}}\times ...
{\partial_{s_{D-1}} k_{\mu_{D-1}}}\times k_{\mu_D}$.

It should be noticed that our $D>1$-dimensional
$\rho_s$ is intrinsically nonlinear in terms of the primary variable $\phi$.
Consequently, the algebra of density operators $\rho_s$ differs from
the U(1) Kac-Moody algebra of $a_{\vec n}({\vec q})$ \cite{HM}
by terms containing higher orders gradients
\bea
[\rho_s ({\vec r}),\rho_{s^\prime}({\vec r}^{\prime})]=
\{1+{1\over 2|\partial_s {\vec k}_{F_{0}}|}
({\vec n}\times{\vec \nabla})(\partial_s \phi -\phi\partial_s )+\\
\nonumber
+{1\over 4|\partial_s {\vec k}_{F_{0}}|^2}
((\partial_s {\vec \nabla}\phi)^2 +({\vec \nabla}\phi)^2\partial^{2}_{s}
-2(\partial_s {\vec \nabla}\phi {\vec \nabla}\phi)  \partial_{s})
\}
\delta({\hat s}-{\hat s}^{\prime}){i({\vec n}{\vec \nabla})\over  2\pi}
\delta({\vec r}-{\vec r}^{\prime})
\eea
The equation of motion for ${\vec k}_F ({\vec r}, t, {\hat s})$
in presence of external electric ${\vec E}
=-\partial_t {\vec A}-{\vec \nabla}A_0$ and magnetic $B={\vec \nabla}
\times{\vec A}$ fields
 can be derived as the  Euler-Lagrange
equation for the Lagrangian  (2.6)
\be
\partial_t w=\{H,w\}_{MB}
\ee
within the approximation (2.10). In the lowest order in $\hbar$ (2.15)
 reads as the standard kinetic equation
(in what follows we shall drop the subscript in the notation
of ${\vec k}_F$):
\be
\partial_t {\vec k}=({\vec v}{\vec \nabla}){\vec k}+{\vec E}+\times{\vec v} B
\ee
To stress a parallel with
a single particle equation of motion we  introduced in (2.16) a generalised
 "Fermi velocity" $\vec v$  defined in terms of a
second  variation of the Hamiltonian $H$ with respect to $\delta{\vec k}$:
\be
\epsilon_s ={\delta H\over \delta\rho_s},
{}~~~~~~~~~~~~~ \delta\epsilon_s  ={\vec v}_s\delta{\vec k}_s
\ee
To complete the scheme we also present
 a gauge invariant current
\be
{\vec j}={\delta H\over \delta{\vec A}}
=\times\int d{\vec r}^{\prime}
\oint{ds^{\prime}\over 2\pi}f_{ss^\prime}({\vec r}-{\vec r}^{\prime})
\partial_{s^\prime}
{\vec k}_{s^\prime}({\vec r}^{\prime})
\ee
satisfying the continuity
equation $(\partial_t\rho+{\vec\nabla}{\vec J}=0)$
provided the equation of motion (2.16) is fulfilled. We stress that
the definition (2.18) which involves spatial gradient rather than a time
derivative follows from the chiral property of the bosonic field $\phi$.
Actually, the relation (2.18) is a familiar one in the conventional Landau-
Fermi liquid theory.

Being estimated on functions $w({\vec p},{\vec q})$ given by (2.10)
the Lagrangian (2.6) acquires the form
\be
L=\int d{\vec r}{\oint}{ds\over (2\pi)^2}
({1\over 2}D_0 \phi ({\partial_s{\vec k}}\times
{\vec k}) +{1\over 4}{\partial_s \phi}
({\vec k}\times{\vec E}-D_0\phi B))-H
\ee
One can also check that
the kinetic equation (2.16) follows from (2.19)
 as the Euler-Lagrange equation
if one understands a local density
variation  as
$\delta\rho_s ={1\over 2\pi}{\partial_s {\vec k}}\times
\delta{\vec k}$ according to its geometrical interpretation.

Following Ref.\cite{H} we choose a simple
form of the Hamiltonian
which is quadratic in $\rho_s$
\be
H={1\over 2}\int d{\vec r}
\oint{ds\over 2\pi}\int d{\vec r}^{\prime}
\oint{ds^{\prime}\over 2\pi}f_{ss^{\prime}}
({\vec r}-{\vec r}^{\prime})\rho_s (\vec r)
\rho_{s^{\prime}} (\vec r^{\prime})
\ee
where the diagonal part of the quadratic form $f_{ss^{\prime}}
({\vec r}-{\vec r}^{\prime})=
v_F \delta(s-s^{\prime})\delta({\vec r}-{\vec r}^{\prime})
+\Gamma_{ss^{\prime}}({\vec r}-{\vec r}^{\prime})$
includes a bare kinetic
energy of fermions near the Fermi surface.

An important  feature of the Lagrangian (2.19) is that it remains essentially
nongaussian in terms of the fundamental field $\phi$
 even in absence of interactions
$(\Gamma_{ss^{\prime}}=0)$. Thus the theory (2.19) is
"geometrical" in the same sense as, say, the nonlinear
$\sigma$-model is.

Although nongaussian terms in (2.19) contain extra gradients they are not
necessarily negligible even in the longwavelength limit if the relation
$q^{2}_t \gsim k_F q_n$ between tangential and normal components of a
typical transferred momentum $\vec q$ holds.
In   accordance with a previous discussion an appearance of these terms
reflects a finite Fermi surface curvature.
Being combined with quadratic terms
they are supposed to reproduce effects of the collision integral which
is introduced
in the phenomenological Landau theory to account quasiparticle scattering.

Strictly speaking,  the most
 complete account of the Fermi surface curvature also requires
to add to the Hamiltonian (2.20)
terms $\sim v_F \rho_{s}^3$ to represent a kinetic
energy of fermions with parabolic dispersion.
Here it is worth to remind an example
of the collective field theory of 1D
free nonrelativistic fermions
which was exactly mapped onto the cubic
bosonic Hamiltonian
$H={1\over 2}(\rho_{R}^2 +\rho_{L}^2 )+
{1\over 6k_F}(\rho_{R}^3 +\rho_{L}^3 )$
\cite{AJ}.
The present formalism is well suited to accommodate those cubic terms too.

Also the Hamiltonian (2.20) has  to be  improved in one intends to include
scattering processes corresponding to the Cooper channel \cite{HM}.
However, it is believed that the form (2.20) could be sufficient
to study an interesting case of a "strange metal"
governed by interactions singular
at small scattering  angles.

Another feature of the Lagrangian (2.19)
specific for two dimensions is an appearance
of the famous Chern-Simons structure ${\sigma_{xy}\over 2}
AdA$ where $AdA={\vec A}\times {\vec E}-A_0 B$
with the coefficient given by the circulation (first Chern class)
\be
\sigma_{xy}={1\over 2}\oint {\partial_s
\phi}{ds\over (2\pi)^2}
\ee
It seems natural to assume that in the case of a zero external magnetic
field the winding number
(2.21)  must vanish so the
Lagrangian (2.19) does not contain parity
odd terms.

On the contrary, one could interpret
the case of the "twisted" Fermi surface
characterised by the lowest nontrivial value of circulation
$\oint {\partial_s \phi}{ds\over 2\pi}
=1$ giving $\sigma_{xy}
=1/4\pi$ (in absolute units of ${e^2\over \hbar}$)
as a proper effective
description of the metal-like state at half filled lowest Landau
level. We shall further comment on this point in the section 4.

\section{Density response function}

The formalism of the preceding section  enables a calculation of
(gauge invariant) correlation functions such as the density response function
without calculating first
(gauge non-invariant) one-particle Green functions.

Keeping in the Lagrangian (2.19) only terms quadratic in $\phi$ one
encounters the problem of a diagonalization of a quadratic form.
Then one obtains a density response function
\be
K_0 (\omega, {\vec q})=<\rho(\omega, {\vec q})\rho(-\omega, -{\vec q})>=
\oint{ds\over 2\pi}\oint{ds^{\prime}\over 2\pi}
({\vec n}_s {\vec q})({\vec n}_{s^{\prime}} {\vec q})
G^{0}_{ss^{\prime}}(\omega, {\vec q})
\ee
where the correlator $G_{ss^{\prime}}(\omega, {\vec q})=
<\phi_s (\omega, {\vec q})\phi_{s^{\prime}}(-\omega, -{\vec q})>$
is given by the
inverse of the quadratic form
\be
G^{0}_{ss^{\prime}}(\omega, {\vec q})=[(({\vec n}_s {\vec q})\omega-v_F
({\vec n}_s {\vec q})({\vec n}_{s^{\prime}} {\vec q}))\delta_{ss^{\prime}}-
\Gamma_{ss^{\prime}}(q)]^{-1}
\ee
In  absence of interactions and in the case of a circular Fermi surface
(3.2) becomes diagonal in $s$-space
and (3.1)
amounts to the longwavelength approximation for the free fermion buble
\be
K_0 (\omega, {\vec q})
=\Pi_0 (\omega, {\vec q})=
\oint{ds\over 2\pi}
{({\vec n}_s {\vec q})\over {\omega- v_F({\vec n}_s {\vec q})}}
\ee
In the case
of a rotationally invariant
interaction $\Gamma_{ss^{\prime}}(q)=V(q)$
(3.2) reproduces the results of RPA. To see that one can simply
expand the inversed operator
 into a series in powers of $\Gamma_{ss^{\prime}}(q)$
which yields
\be
K_{RPA}(\omega, {\vec q})
={V(q)\over {1+\Pi_0(\omega, {\vec q}) V(q)}}
\ee
where $\Pi_0(\omega, {\vec q})$ is given by (3.3).
It is well known that
in the case of long-ranged
interactions the RPA compressibility
$K(0, {q\rightarrow 0})\sim V^{-1}(q)$
vanishes at $q\rightarrow 0$.

Notice that in the gaussian approximation equivalent to RPA
the spectrum of the collective mode $\omega\sim
q{V^{1/2}(q)}$ lying
outside of the particle-hole continuum remains undamped at zero temperature.
However this property
of collective excitations can only hold in the case
of 1D Luttinger liquid  where dynamics of
low-energy bosonic
density modes is governed by an exactly quadratic Lagrangian \cite{DL}.

To proceed beyond RPA one has to consider
the nonlinear equation of motion (2.16) written in terms of $\phi (x)$
\be
\partial_t \phi_s ({ x})+\oint {ds^{\prime}\over (2\pi)}
f_{ss^{\prime}}({x}-{x}^{\prime})({\vec n}_{s^{\prime}}+
{1\over 2}{\vec \nabla}\partial_{s^\prime}\phi_{s^{\prime}}({x}^{\prime})
\times){\vec \nabla}\phi_{s^{\prime}}({x}^{\prime})=0
\ee
To compute
the response function $G_{ss^{\prime}}$ corresponding
to the equation (3.5) we apply a nonperturbative
 eikonal-type method similar to the one used in the context of
the Navier-Stokes equation \cite{F}.
Similar to the case of the one-particle fermion Green
function studied in \cite{KS}
the use of the eikonal method becomes possible due to the presence
of a large term in (3.5) which contains a bare Fermi velocity $v_F$.
Analogously to the case of an advection of a passive scalar  in the
theory of turbulence \cite{turb}
the method provides a consistent summation of
infrared relevant terms in a perturbation theory for the Lagrangian (2.19).

Using (3.5) one arrives at the equation for the Fourier transform
$G(q,x)=\int (dq)e^{iq(x-x^{\prime})}\\
G(x, x^{\prime})$ of the (translationally non-invariant)
response function in an external field $\phi(x)$
\be
({\vec n}_{s}{\vec \nabla})[\partial_t
G_{ss^{\prime}}(q,x)-\int (dp)\oint{ds^{\prime\prime}\over (2\pi)}
f_{ss^{\prime\prime}}({\vec q}-{\vec p})
e^{ipx}(\delta({p}){\vec n}_{s^{\prime\prime}}+
{1\over 2}{\vec p}\partial_{s^\prime\prime}\phi_{s^{\prime\prime}}
(p)\times)
G_{s^{\prime\prime}s^{\prime}}(q,x)]=\delta_{ss^{\prime}}
\ee
 According to \cite{F}
the response function can be searched in the Fradkin's integral form
\be
G_{ss^{\prime}}(q,x)=
i\int_0^{\infty}d\nu \oint {ds^{\prime\prime}\over 2\pi}
<s|e^{i\nu G_{0}^{-1}(q)}|s^{\prime\prime}>
<s^{\prime\prime}|\exp(i\Psi(q,x))s^{\prime}>
\ee
where $G^{0}_{ss^{\prime\prime}}$ is given by (3.2).

Then keeping the term of the lowest order in $\vec q$ one obtains
\be
<s|\Psi(q,x)|s^{\prime}>=i\int_0^{\nu}d\nu^{\prime}\int (dk)
e^{ikx}
\oint {ds^{\prime\prime}\over 2\pi}
f_{ss^{\prime\prime}}(q-k)({\vec k}\times{\vec q})
\partial_{s
^{\prime\prime}}\phi_{s^{\prime\prime}}(k)
<s^{\prime\prime}|e^{i\nu^{\prime} G_{0}^{-1}(k)}|s^{\prime}>
\ee
where $(dk)={d\Omega d^2 {\vec k}\over (2\pi)^3}$.
All terms of higher orders in $\vec q$ can be found by recursion \cite{KS}.

The consistency condition  requires to average (3.7) over fluctuations of
$\phi(x)$. Then  for the response
function\\
$
G_{ss^{\prime}}(q)=<\phi_s (q)\phi_{s^{\prime}} (-q)>=
\int {\cal D}\phi  G(q|\phi(x)) e^{{i\over 2}\int\phi K\phi}
$\\
we get the following equation
\bea
G_{ss^{\prime}}(q)
=i\int_{0}^{\infty}d\nu <s|e^{i\nu G_{0}^{-1}(q)}|s^{\prime}>
\exp({i}\int(dk)\oint_{s_1} ({\vec q}{\vec n}_s )\oint_{s_2}
({\vec q}{\vec n}_{s^{\prime}})
({\vec k}\times{\vec q})^2 \\
\nonumber
f_{ss_{1}}(q-k)f_{s^{\prime}s_{2}}(k-q)
\partial_{s_1}\partial_{s_2}G_{s_1 s_2}(k)
<s_1|[(e^{i\nu^{\prime}G_{0}^{-1}(k)}-1)G_{0}^{2}(k)-
i\nu^{\prime}G_{0}(k)]|s_2>)
\eea
This equation can be treated using various approximations which are supposed
to give corrections to the RPA expression (3.2). In particular,
one could expect to find this way a damping of collective excitations
absent in the gaussian theory. The simplest
 approximation would be to substitute $ G_{s_1 s_2}(k)$
in the exponent
in (3.9) by the formula (3.2).

It's also worth to compare the equation
(3.9)
with the eikonal formula  for $K(\omega,{\vec q})$ obtained in the original
fermion representation \cite{KS}
\bea
K(\omega,{\vec q})=\int d^{2}{\vec p} n_{\vec p}^{(0)} \int_{0}^{\infty}d\nu
e^{i\nu(\omega +\xi_{\vec p}-\xi_{{\vec p}+{\vec q}}+i\delta)}
\exp(i\int (dk)
V( k)[(1-n^{(0)}_{{\vec p}+{\vec q}+{\vec k}})
\\ \nonumber
{1-e^{i\nu
(\xi_{{\vec p}+{\vec q}}-\xi_{{\vec p}+{\vec q}+{\vec k}}+
\Omega)}
\over {\xi_{{\vec p}+{\vec q}}-\xi_{{\vec p}+{\vec q}+{\vec k}}+
\Omega}}
-n^{(0)}_{{\vec p}+{\vec k}}
{1-e^{i\nu
(\xi_{{\vec p}}-\xi_{{\vec p}+{\vec k}}+\Omega)}
\over {\xi_{\vec p}-\xi_{{\vec p}+{\vec k}}+\Omega}}]^{2})
+(\omega\rightarrow -\omega)
\eea
where $\xi(p)$ is a (in general, nonlinear) fermion dispersion.
An obvious advantage of this formula as compared to (3.9)
is that nonlinear terms in fermion
dispersion are already accounted in (3.10). However, to see the
effect of Ward identities which guarantee a cancelation of self-energy versus
vertex corrections in  (3.10)
one has to treat carefully a combination of
three terms forming a complete square in the exponential factor.
On the other hand, it automatically follows from an appearance
of extra powers of $q$ in the corresponding exponential factor in (3.9)
as it should be in a consistent
hydrodynamics  of interacting fermions.

\section{ Towards
(Non) Fermi liquid theory of the half filled Landau level}
There exists a whole
bunch of experimental evidencies in favor of the compressible
metal-like state at $\nu=1/2$ \cite{exp}. Experiments on geometric resonance
in the antidot array as well as magnetic focusing show convincingly
that quasiparticle excitations at half filling experience no magnetic field.

A theoretical explanation of the phenomenon proposed in \cite{HLR}
was based on the Jain's idea of attaching a pair of fictitious
flux quanta to each electron to compensate
an external magnetic field
\cite{J}. Moreover it was suggested to treat FQHE states belonging to the
sequence of fractions $\nu={N\over {2N+1}}$ converging towards $\nu=1/2$
as IQHE states of quasiparticles occupying $N$ Landau levels in the net
field $B_{eff}={1\over 2N+1}B_{ext}$.

In the framework of this picture
the oscillating magnetoresistivity $\Delta
R_{xx}(B,T)$ at $\nu\neq 1/2$, for example,
can be treated as Shubnikov-de Haas oscillations
in the system of quasiparticles
with some effective mass $m^{*}$ placed into the field $B_{eff}$.
In the physically relevant case of unscreened Coulomb interaction
the theory \cite{HLR} predicts, in particular, a weak (logarithmic)
divergence of $m^{*}\sim \ln|\delta\nu|$
as $\delta\nu=|\nu-1/2|$ approaches zero.

An extensive study of the effects of the transverse
statistical gauge interaction
carried out in \cite{AIM} led to the conclusion that in many respects
the system looks like Fermi liquid except for the "marginal" change
$\epsilon\rightarrow \Sigma(\epsilon)\sim \epsilon
\ln\epsilon$ in its one-particle Green function.
Another new feature is an extremely
 weak divergency of the $2k_F$ scattering
amplitude \cite{AIM} which presumably does not lead to a divergency of
any physical polarizability.

These results imply that
in spite of a strong lowest order renormalization
of the (gauge non-invariant) fermion propagator
transverse gauge fluctuations  have only a little effect
on physical observables. In addition to the results of
the self-consistent diagrammatic approach of \cite{AIM} this conclusion
was confirmed by a straightforward  two-loop calculation of irreducible
density and current polarizabilities  which showed no sign of a divergent
effective mass either \cite{MIT}.

On the other hand, as opposed to
earlier reported results,
the most recent measurements of magnetoresistvity
at relatively
high temperatures on
both electron-like \cite{exp1} and hole-like \cite{exp2}
systems revealed a strong dependence of $m^{*}$ on $\nu$
extracted from the Fermi liquid formula for the first harmonics of SdH
oscillations
$$\Delta
R_{xx}(B,T)\sim {Tm^* /B_{eff}
\over {sh (Tm^* /B_{eff})}} e^{-m^* /B_{eff}\tau}$$
Namely, the results obtained in \cite{exp2} were best fitted by
the function
$$
m^{*}(\delta\nu) \sim \exp |\delta\nu|^{-3/2}
$$
while the authors of \cite{exp1} found a rather strong
power-law behavior of $m^{*}(\delta\nu)$.

When contrasting these experimental observations  against the
theoretical conclusions \cite{HLR}, \cite{AIM}
it seems that a strong dependence  of the effective mass $m^{*}
(\delta\nu)$
does not find an immediate explanation
in the framework of the  "marginal" Fermi liquid behavior.
Although one should be cautious about applying Fermi liquid formulae
to the analysis of data obtained in \cite{exp1},\cite{exp2}
(see, however, \cite{P}) it nevertheless gives enough motivation to search for
an alternative description which uses no spurious gauge field at all
(or, equivalently, the statistical gauge field  is integrated out
exactly) for an independent check of predictions of the gauge theory.
It is quite likely that a possible  candidate could be a sort of
"nonlinear Landau-Fermi liquid
 theory" which one can use to understand
a drastic effective mass dependence on $\nu$ (for a related discussion see
\cite{SH}).

To make an attempt in this direction we propose the Lagrangian
describing the $\nu=1/2$ state
in the form  (2.19)
where  $A_\mu$ is now a sum of the
statistical gauge field $a_\mu$ and the external electromagnetic potential
${\vec A}_{ext}={B_{ext}\over 2}(y,-x)$.
As we mentioned above it requires the field $\phi$
to have a nonzero winding number
$\oint {ds\over (2\pi)^2} {\partial_s \phi}=1$.

Notice that in comparison with the conventional Landau-Ginsburg-type
description of the odd-denominator FQHE states given in terms
of the phase field $\phi({\vec r},t)$ the conjectured
theory of the even-denominator state involves an extra
variable $s$ along the boundary of an extended region in momentum space
identified with the bare Fermi surface of transformed ("neutral") fermions.

Varying the Lagrangian
(2.19) with respect to $a_0$ one obtains an intrinsic  operator
 relation between local fermion density and statistical flux
\be
{1\over (2\pi)^2}{\oint}ds({\partial_s \phi} \nabla\times{\vec a}-
{\partial_s{\vec k}}\times
{\vec k})=0
\ee
Due to the local constraint (4.1)
the external field ${\vec A}_{ext}$ is
canceled out by the averaged flux and
the bare Lagrangian of gauge field fluctuations
is given by the Chern-Simons term and by the pairwise interaction
$V({\vec r}-{\vec r}^{\prime})$
rewritten in terms
of ${\vec a}$
\be
L_g ={1\over 2}\int d{\vec r}d{\vec r}^{\prime}
({\vec \nabla}\times{\vec a})_{{\vec r}}
V({\vec r}-{\vec r}^{\prime})
({\vec \nabla}\times{\vec a})_{{\vec r}^{\prime}}
+{1\over 8\pi}\epsilon_{\mu\nu\lambda}a_{\mu}\partial_{\nu}a_{\lambda}
\ee
Intending to deal with gauge invariant quantities
one can choose the gauge ${\vec \nabla}\vec a=0$ and
then  integrate $a_\mu$ out to
end up with the effective Lagrangian written solely in terms of densities
\be
L_{eff}={\oint}{ds\over 4\pi^2}\rho_s{\partial_t \phi_s}
-
{1\over 2}
{\oint}{ds\over 2\pi}{\oint}{ds^\prime\over 2\pi}
(\rho_{s}({\vec r})V({\vec r}-{\vec r}^{\prime})\rho_{s^{\prime}}
({\vec r}^{\prime})+i
{\vec \nabla}\times {\vec j}_{s}({\vec r})
<{\vec r}|{1\over {\vec \nabla}^2}|{\vec r}^{\prime}>
\rho_{s^{\prime}}({\vec r}^{\prime}))
\ee
where $\rho_s$ and $\vec j_s$ are given by (2.13) and (2.18) respectively and
the induced $\rho-j$ coupling is due to the
statistical Chern-Simons
interaction contained  in (2.19).

It's worth mentioning here a possibility of
an independent microscopic check of
validity of the effective theory (4.3).
Namely, at $\nu=1/2$  one can choose to work with a basis
of coherent states of $\cal N$ fermions on the lowest Landau level
labeled by a set of $\cal N$ two-dimensional momenta $\{{\vec k}_i\}$
\cite{H1}:
\be
|\{{\vec k}\}>
=det e^{i{\vec k}_i {\vec R}_j}\prod_{i<j}
 (z_i -z_j)^2 e^{-{1\over 4}|z_i|^2}
\ee
where $\vec R_i$ is the center of Larmor's orbit operator of $i^{th}$
electron.
This choice of the basis
is not accidental. The conventional (in our case symmetrical)
Jastrow factor in (4.4)
takes care of short range correlations and provides a good
variational energy. Due to the antisymmetry of the entire
wave function all ${\vec k}_i$ have to be
distinct, therefore the other factor is the Slater determinant
of $\exp (i{\vec R}_i {\vec k}_j)$ which reflects an alleged
metal-like behavior governed by long-range correlations.
Then using a kind of the collective field approximation \cite{IKS}
one can describe different patterns of occupied ${\vec k}_j$-states in terms
of the Fermi momentum tracing the boundary of the filled region in
${\vec k}$-space \cite{H1}.

However the basis of coherent states (4.4) is not, of course, orthonormal.
One might expect that it is the overlap between different
states which  causes
the induced $\rho-j$ coupling in the
effective Lagrangian
\be
L={<\{\vec k\}|(i\partial_t -H|\{\vec k\}>\over <\{\vec k\}|\{\vec k\}>}
\ee
An explicit calculation of (4.5) which will provide a
 decisive check of the status of (4.3) remains to be done.

{}From (4.3) one reads off the bare interaction kernel in the form
\be
\Gamma_{ss^\prime}(q)=V(q)+{v_F  \over q^2}
{\vec q}\times ({\vec n}_s -{\vec n}_{s^{\prime}})
\ee
Similar to the case of  pure density-density interactions,
a diagonalization procedure applied to the quadratic form with
$\Gamma_{ss^\prime}$
given by  (4.6) leads to the RPA  density and current response
functions. In particular, (3.2) now yields
\be
K_{RPA}(\omega,{\vec q})
={q^2\Pi_{0}(\omega,{\vec q})
\over {k_F\Pi_{0}(\omega,{\vec q})
(k_F\Pi_{\perp}(\omega,{\vec q})+V(q)q^2)+q^2}}
\ee
where $\Pi_0 (\omega,{\vec q})$ is given by (3.3) and
the longwavelength approximation for the current polarization operator
is
\be
\Pi_{\perp}(\omega, {\vec q})=
v_{F}^{2}\oint{ds\over 2\pi}
{({\vec n}_s {\vec q})
({\vec n}_s \times{\vec q})^2
\over {q^2 (\omega- v_F({\vec n}_s {\vec q}))}}-{1\over 2}v_{F}
\ee
Note that the crossed current-density polarization operator
$\Pi_{H}(\omega, {\vec q})\sim \oint{ds\over 2\pi}
{({\vec n}_s {\vec q})
({\vec n}_s \times{\vec q})
\over {q^2 (\omega- v_F({\vec n}_s {\vec q}))}}$
vanishes in this approximation.

Using the well known asymptotics of $\Pi_{0,\perp}(\omega, {\vec q})$
at $\omega <<v_F q$
\be
\Pi_0 (\omega,{\vec q})
={1\over 2\pi v_F}(1 +i{\omega\over {v_F q}})~~~~~~~~~~~~~
\Pi_\perp (\omega,{\vec q})
={v_{F}\over {2\pi}}
({q^{2}\over k_{F}^{2}} +i{\omega\over v_F q})
\ee
one readily reproduces the RPA compressibility
$K(0,{\vec q})\rightarrow 0\sim 1/V(q)$ and the location of the pole of (4.7)
for small $q$  at
$\omega\sim iV(q)q^3$  in agreement with \cite{HLR}.

Notice that similar results can be  obtained from a direct solution of the
eigenvalue problem in the case of the induced current-current interaction
\cite{KS}
\be
({\vec n}_s {\vec q})(\omega- v_F({\vec n}_s {\vec q}))\phi_s =
{\Gamma_{\perp}(q)\over q^2}
({\vec n}_s \times{\vec q})
\oint {ds^{\prime}
\over 2\pi}({\vec n}_{s^{\prime}} \times{\vec q})\phi_{s^{\prime}}
\ee
where $\Gamma_{\perp}(\omega,{\vec q})=
<({\vec \nabla}\times{\vec a})({\vec \nabla}\times{\vec a})>=
(4\pi)^2 K_{RPA}(\omega,{\vec q})$.

We plan to undertake an analysis of corrections to RPA using the
eikonal-type formula
(3.9) in a separate paper. Although we don't expect that at small
$\omega, q$ the corrections to (4.7)
alter the behavior found in RPA, one might think that this approximation
becomes insufficient when calculating the $2k_F$-response (particularly, if
the RPA result contains a possible divergency \cite{AIM}).

One remark is in order here.
In the case of a zero $q$ but
finite $\omega$ the function $K(\omega, 0)$
has a pole located
at renormalized
cyclotron frequency $\omega_{c}^{*} =B/m^*$ while
according to the Kohn theorem it would have to
occur at the bare one. It was proposed in \cite{M}
 to improve this point by adding an extra
Fermi-liquid interaction $f_{ss^\prime}\sim F_1 ({\vec n}_s{\vec n}
_{s^\prime})$
to the Chern-Simons gauge theory
which  allows to restore the bare mass
${1\over m_0}={1\over m^*}+F_1$  while doing RPA.
However  it still doesn't seem
to reflect the fact that switching interaction off one completely
eliminates quasiparticle dispersion $(v_F\sim {1\over m^*}\rightarrow  0)$.

Moreover  this recipe may appear to lead to a double counting of
the effects of the original interaction $V(q)$ if one
goes beyond RPA. To this end, it is conceivable that
studing the problem (4.3) at small energies one should
put in (4.6) $v_F$
equal to zero for consistency and to assume that the kernel $
\Gamma_{ss^\prime}$ does have higher angular harmonics in contrast
to a naive $\Gamma^{0}_{ss^\prime}(q)=V(q)$ which is a pure $s$-wave.

\section {Conclusions}
In the present paper we discussed a scheme of $D>1$ bosonization
of interacting fermions in external fields
which is based on a geometrical picture of fluctuating Fermi surface.
The natural
description can be done  in terms of the vector field ${\vec k}_F ({\vec r},
{\hat s})$ tracing the space-time dependent shape of the Fermi surface.
We argue
that it
arises in a longwavelength approximation for the formally exact
bosonization scheme in terms of coherent states on quantum phase space
\cite{ind},\cite{K}. Conceptually,
this sort of description can be viewed as a generalization of
the phenomenological Landau theory which leads to a quantum
hydrodynamics incorporating small angle scattering between quasiparticles
around a non-flat Fermi surface.

In contrast to the constructive approach of Refs. \cite{H},
\cite{HM},\cite{FC} leading to a gaussian bosonic theory
the present
 method gives an intrinsically nonlinear one. It is this property which
makes it
possible to account  the effects of the Fermi surface curvature.

It should be mentioned here that in special cases a bare  Fermi surface
having flat faces may preserve them under renormalization.
For instance, it was
argued in \cite{L2} that a square Fermi surface is stable with respect
to parallel face interactions. However, interactions between adjacent faces
are likely to cause rounding of the square Fermi surface.

The bosonic formalism also allows one
to avoid a subtle problem of an explicit
representation of the fermion operator in terms
of bosons when calculating gauge invariant
response functions.

A simple diagonalization of the quadratic form
while neglecting nonlinear terms gives usual RPA results.
To go
beyond RPA we formulate an eikonal-type procedure leading
to integral equations for the density
response function which can be solved iteratively.
In particular, one might expect to obtain damping of collective modes out of
these equations.

In addition, we apply our approach to the compressible state of the
half filled lowest Landau level and formulate a nonlinear
effective theory in terms
of chiral bosons representing density fluctuations.
In the gaussian approximation
the theory reproduces the results of RPA in the gauge theory of $\nu=1/2$
state \cite{HLR}. We intend to study effects of nonlinear terms on RPA
results elsewhere.

\section{Acknowledgements}
The author is indebted to Prof. F.D.M.Haldane for extremely valuable
discussions and suggestions concerning these and related issues.
He is also grateful to the Aspen Center of Physics where
this work was completed.
This work was supported by the Grant NSF-DMR-922407.

\end{document}